\documentclass[sigconf]{acmart}
\usepackage{booktabs} 
\usepackage{xspace}

\usepackage{subfig}

\usepackage{multirow} 
\usepackage{array} 
\usepackage{tabularx} 
\usepackage{enumitem}

\usepackage{xcolor,colortbl}

\usepackage{float}
\usepackage{placeins}

\graphicspath{{Figures/}}

\newcommand{\ie}{\emph{i.e.,}\xspace}
\newcommand{\eg}{\emph{e.g.,}\xspace}

\newcommand{\paratitle}[1]{\vspace{1ex}\noindent \textbf{#1}}

\newtheorem{observation} {\textbf{Observation}}

\setcopyright{rightsretained}

\acmDOI{}

\acmISBN{}

\acmConference[SIGIR '19]{The 42nd International ACM SIGIR Conference on Research \& Development in Information Retrieval}{July 21--25, 2019}{Paris, France}
\acmYear{2019}
\copyrightyear{}

\acmPrice{15.00}

\begin{document}

\title{A Study on Agreement in PICO Span Annotations}

\author{Grace E. Lee and Aixin Sun}
\orcid{}
\affiliation{%
  \institution{School of Computer Science and Engineering, Nanyang Technological University, Singapore}
  \streetaddress{}
  \city{}
  \state{}
  \postcode{}
}
\email{leee0020@e.ntu.edu.sg;axsun@ntu.edu.sg}

\begin{abstract}
In evidence-based medicine, relevance of medical literature is determined by predefined relevance conditions. The conditions are defined based on PICO elements, namely, \textbf{P}atient, \textbf{I}ntervention, \textbf{C}omparator, and \textbf{O}utcome. Hence, PICO annotations in medical literature are essential for automatic relevant document filtering.
However, defining boundaries of text spans for PICO elements is not straightforward.  
In this paper, we study the agreement of PICO annotations made by multiple human annotators, including both experts and non-experts.
Agreements are estimated by a standard span agreement  (\ie matching both labels and boundaries of text spans), and two types of relaxed span agreement (\ie matching labels without guaranteeing matching boundaries of spans).
Based on the analysis, we report two observations: (i) Boundaries of PICO span annotations by individual human annotators are very diverse.
(ii) Despite the disagreement in span boundaries, general areas of the span annotations are \textit{broadly agreed} by annotators.
Our results suggest that applying a standard agreement alone may undermine the agreement of PICO spans, and adopting both a standard and a relaxed agreements is more suitable for PICO span evaluation.
\end{abstract}


\maketitle

\section{Introduction}
\label{sec:Introduction}

In evidence-based medicine,
it is crucial for medical professionals to effectively access and find relevant literature since medical decisions are made based on primary evidence.
Relevance of a document depends on relevance conditions which are defined using PICO framework: \textbf{P}atient (problem, population), \textbf{I}ntervention, \textbf{C}omparator, and \textbf{O}utcome. 
PICO elements identified in medical literature, therefore, are critical for effective retrieval of medical literature.

Existing datasets used for automatic identification of PICO elements are sentence-level annotations~\cite{PICO_dataset_1000_doc_sent_level,sentence_lvl_PICO_anno_Grace_Chung}. 
Recently, toward more accurate and detailed identification, a large-scale PICO \textit{span} annotation dataset (EBM-PICO) is released~\cite{PICO_dataset}. 
EBM-PICO consists of 5,000 abstracts of medical literature with PICO span annotated by medical experts and non-experts.
Annotated spans can be either a single word or a long phrase.

\begin{figure}
  \includegraphics[width=0.75\linewidth]{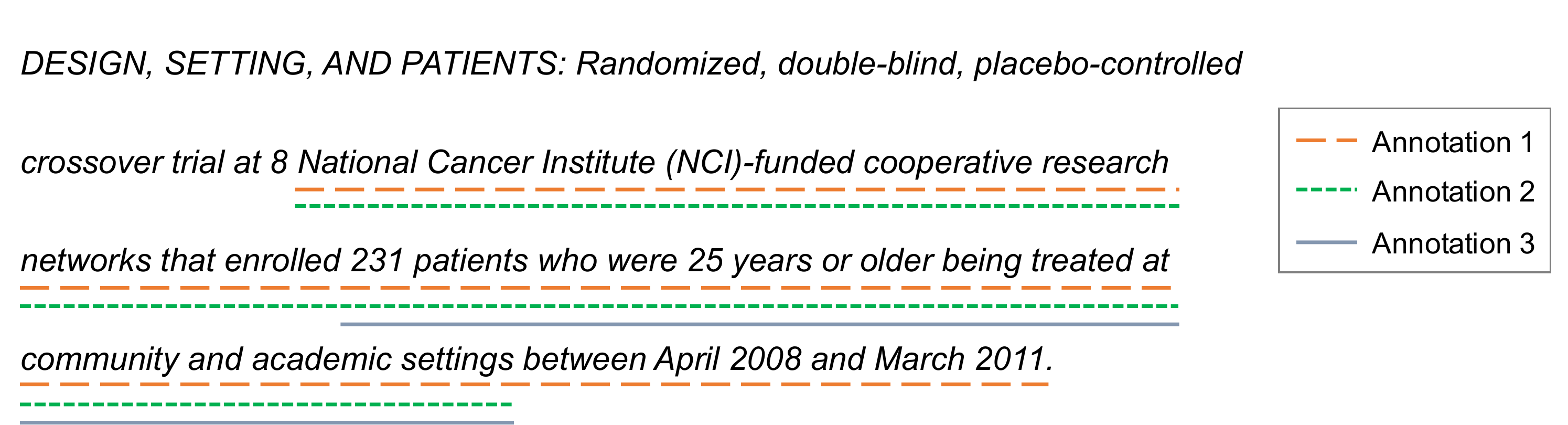}%
\caption{Example of span annotations in EBM-PICO dataset. This sentence (from PMID: 23549581) is annotated by 12 annotators (4 experts and 8 non-experts) and 3 different span annotations are made by them. The 3 spans have the same label \textbf{P}, but different boundaries.}
\label{fig:diff_anns_ex}
\end{figure}

PICO elements are presented in a \textit{descriptive} manner in medical literature.
Verbose descriptions of PICO elements make it nontrivial to decide spans of PICO annotations.
Figure~\ref{fig:diff_anns_ex} shows an example sentence and its annotations made by annotators in EBM-PICO dataset. 
The example sentence contains information about P element.
For a given sentence, the three different spans are annotated and all spans are labeled as \textbf{P} and indicate an acceptable information for P, even though each span has distinct boundaries.
Among them, considering one span annotation correct and the rest incorrect may not be a reasonable decision.

In this paper, we study agreement in PICO span annotations made by different human annotators in  EBM-PICO dataset.
Specifically, we evaluate the annotation agreement using two types of measures: \textbf{exact span agreement} and \textbf{relaxed span agreement}.
Exact span agreement is a standard evaluation approach for text span annotations. It evaluates both boundaries and labels of two spans.
In relaxed span agreement, we analyze PICO annotation in terms of two variants: \textit{one-side boundary} (\textbf{OB}) agreement and \textit{token overlap} (\textbf{TO}) agreement.
OB and TO agreements evaluate whether annotations are with same label but without guaranteeing the exact matching start and end boundaries between two spans.

As the annotations in EBM-PICO dataset are made by both medical experts and non-experts, we estimate the agreement within each of expert and non-expert groups, and across the expert and non-expert groups. 
Our study shows extremely low-level of exact span agreement, but significantly high-level of relaxed span agreements in both within and across groups
The large improvement in relaxed span agreement indicates the general area of annotations is mostly agreed despite unmatched boundaries. 
Our results suggest that applying exact agreement alone may underestimate the agreement of PICO span annotations. Therefore adopting both exact and relaxed agreements is more suitable for PICO span evaluation such as PICO span recognition task. 
Lastly, we present how the two agreements are leveraged in PICO span recognition task.

\section{PICO Span Annotation Dataset}
\label{sec:dataset}
We use  EBM-PICO dataset~\cite{PICO_dataset} which provides PICO \textit{span} annotations on 5,000 medical literature abstracts. Examples of annotations are shown in Figure~\ref{fig:diff_anns_ex}. We study this dataset because of its fine-grained annotations and a large scale. Other existing PICO annotation datasets are sentence-level annotations~\cite{PICO_dataset_1000_doc_sent_level,sentence_lvl_PICO_anno_Grace_Chung} and/or contain only a few hundreds of documents~\cite{non_sent_level_PICO_anno_but_small}. Note that, in EBM-PICO dataset Intervention (I) and Comparator (C) are combined as a single element I, so that each document has three types of annotations:  \textbf{P}, \textbf{I}, and \textbf{O}.
Each annotation (\ie a text span as in Figure~\ref{fig:diff_anns_ex}) has one of the three labels.

The EBM-PICO dataset provides annotations by individual annotators, and also aggregated annotations which combine annotations by individuals (used in Section~\ref{sec:PICORecognition}).
We focus on the individuals' annotations to study agreements of PICO span annotations.
Individual annotators include two groups: Amazon Mechanical Turk (MTurk) workers as non-experts and medical experts.
Specifically, all the 5,000 documents have annotations of PIO elements by MTurk workers. For each document, annotation process is conducted by at least three MTurk workers. 
Among the 5,000 documents, 200 documents have annotations done by medical experts.\footnote{\scriptsize We note that a very small number of documents in the dataset have no annotations.}
Each of the 200 documents has at least two experts' annotations. Hence, these 200 documents have annotations by both MTurk workers and medical experts.

\section{PICO Span Agreement}
\label{sec:span_agreement}

We measure agreement of PICO span annotations by computing $F1$.
Given a pair of annotators, $A$ and $B$, for a document, annotations made by $A$ are first considered as \textit{gold standard annotations}, and annotations by  $B$ are considered as \textit{predicted annotations}. We also switch $A$ and $B$ to calculate $F1$. Similar evaluation scheme has been used in~\cite{yedda}. As a document has more than two annotators in EBM-PICO dataset, we average these values for all pairs of annotators, depending on the evaluation scenario (\eg within or cross non-expert and expert group evaluations). $F1$ is estimated by two kinds of agreement definitions.

\begin{table}
\centering
\caption{Five examples of predicted annotations (Predicted) and their evaluations by different agreements (\ie Exact, OB, TO). The example sentence consists of 7 tokens, and tokens in blue denote annotated text span (the same label is assumed in all annotations).
o/x indicates a correct/wrong prediction against gold annotation (Gold) by a given agreement.}
\label{tab:agreement_correct_examples}
\begin{tabular}{c|c|c|c|c|c}
\toprule
\textbf{No.} &\textbf{Predicted}&\textbf{Gold}&  \textbf{Exact} & \textbf{OB}& \textbf{TO}\\
\midrule
1&\includegraphics[width=0.1\textwidth]{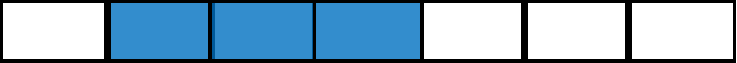} &\multirow{5}{*}{\includegraphics[width=0.1\textwidth]{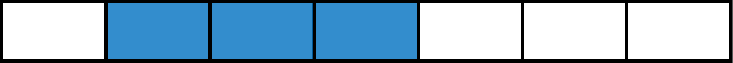}}& o& o& o \\
2&\includegraphics[width=0.1\textwidth]{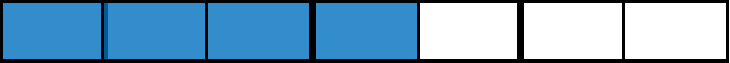} & & x& o& o\\
3&\includegraphics[width=0.1\textwidth]{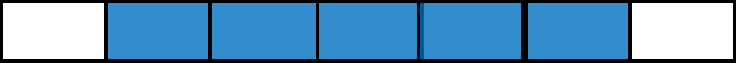} & & x& o& o\\
4&\includegraphics[width=0.1\textwidth]{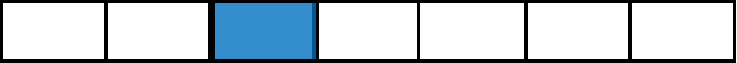} & & x& x& o\\
5&\includegraphics[width=0.1\textwidth]{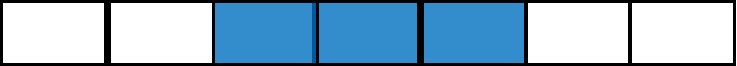} & & x& x& o\\
\bottomrule
\end{tabular}
\end{table}

\paratitle{Exact and Relaxed Span Agreements.}
In exact span agreement, two annotations agree with each other, if both have the same label and the same boundaries of text spans.
By comparison, we report two types of relaxed span agreement: one-side boundary (OB) agreement and token overlap (TO) agreement.
Both OB agreement and TO agreement estimate the agreement in terms of whether two annotations indicate a text span with same label, but allowing different boundaries.
Table~\ref{tab:agreement_correct_examples} shows 5 examples of predicted annotations and their evaluations under the
different span agreement definitions.
The example sentence has 7 tokens and each cell represents a token. The cells in blue indicate a text span annotation. All examples shown in Table~\ref{tab:agreement_correct_examples} have the same label.

In \underline{o}ne-side \underline{b}oundary (\textbf{OB}) agreement, if either side of boundaries in predicted annotation is matched with a corresponding boundary of gold standard annotation, and there is at least one overlapped token between the two span annotations, then the predicted annotation is considered as a correct prediction.
In other words, a predicted span annotation can be larger or smaller than a gold standard span annotation, but the two spans share at least the same start or end boundaries.
For example, in Table~\ref{tab:agreement_correct_examples}, No. 1, 2, and 3 predicted annotations are correct annotations in OB agreement.

\underline{T}oken \underline{o}verlap (\textbf{TO}) agreement is defined with a more relaxed setup than one-side boundary (OB) agreement.
In TO agreement, without considering boundaries of span annotations, if a predicted span annotation has at least one overlapped token with a gold standard span annotation, TO agreement counts the predicted annotation as a correct prediction. Table~\ref{tab:agreement_correct_examples}, all five predicted annotations are correct under the TO agreement because of overlapped tokens.

Note that, the original EBM-PICO dataset paper reports token-level annotation agreement within expert annotators~\cite{PICO_dataset}. In their evaluation, token-level agreement is defined on individual tokens and does not consider span-level agreement.\footnote{\scriptsize Token-level agreement has more evaluation instances than span-level agreement (\ie Exact, One-side boundary (OB), Token overlap (TO)) since a span consists of several tokens.} In our evaluation, both exact span agreement and the two relaxed span agreements are at span-level.

\begin{table*}
\centering
\caption{Average (standard deviation) $F1$ scores estimated within MTurk workers on 5000 documents (MTurk-5000). For the 200 documents containing annotations by both MTurk workers and medical experts, the within group agreement of MTurk workers (MTurk-200) and medical experts (Expert-200) are reported. The $F1$ scores are computed based on different agreements.}
\label{tab:exact_and_relaxed_span_agree}
\begin{tabular}{c|ccc||ccc|ccc}
\toprule
& \multicolumn{3}{c||}{\textbf{Exact span agreement}} & \multicolumn{6}{c}{\textbf{Relaxed span agreement, Avg. $F1$ Score (Standard deviation)} }\\
& \multicolumn{3}{c||}{\textbf{Avg. $F1$ Score (Standard deviation)}} & \multicolumn{3}{c}{\textbf{One-side Boundary (OB)}}& \multicolumn{3}{c}{\textbf{Token Overlap (TO)}} \\
\midrule
 & MTurk-5000 & MTurk-200 & Expert-200 & MTurk-5000 & MTurk-200 & Expert-200 & MTurk-5000 & MTurk-200 & Expert-200 \\
\midrule
P &  0.187 (0.136)&  0.202 (0.125)  &  0.395 (0.201) & 0.361 (0.179)& 0.385 (0.157)  &  0.680 (0.200) & 0.421 (0.190)&  0.441 (0.161)  &0.737 (0.189)\\
I &  0.093 (0.085)& 0.137 (0.095)  &0.576 (0.301)  &  0.187 (0.112) & 0.235 (0.120) &0.732 (0.271)  & 0.241 (0.123)& 0.282 (0.128)  &0.758 (0.269)\\
O &  0.064 (0.053)&  0.078 (0.054)  & 0.357 (0.167) &  0.139 (0.089) & 0.175 (0.093)  &0.654 (0.178) & 0.215 (0.121)&  0.256 (0.115) &0.713 (0.169)\\
\bottomrule
\end{tabular}
\end{table*}

\begin{table*}
\centering
\caption{Precision, Recall, and $F1$ values of cross-group annotation agreement between MTurk workers (predicted annotation) and medical experts (gold standard). The average values (standard deviation in parenthesis) of 200 documents are reported.}
\label{tab:inter_group_agreement}
\begin{tabular}{c|ccc||ccc|ccc}
\toprule
& \multicolumn{3}{c||}{\textbf{Exact}} & \multicolumn{6}{c}{\textbf{Relaxed span agreement}}\\
& \multicolumn{3}{c||}{\textbf{span agreement}} & \multicolumn{3}{c}{\textbf{One-side Boundary (OB)}}& \multicolumn{3}{c}{\textbf{Token Overlap (TO)}} \\
\midrule
 &Pre &Rec &$F1$ & Pre &Rec &$F1$ &Pre &Rec &$F1$\\
\midrule
P     & 0.266 (0.126) & 0.338 (0.144) & 0.275 (0.125) & 0.483 (0.144) & 0.624 (0.172) & 0.496 (0.144) & 0.537 (0.146) & 0.707 (0.178) & 0.553 (0.143) \\
I     & 0.243 (0.131) & 0.332 (0.186) & 0.257 (0.141) & 0.360 (0.149) & 0.570 (0.239) & 0.391 (0.159) & 0.408 (0.154) & 0.769 (0.364) & 0.457 (0.173) \\
O     & 0.147 (0.078) & 0.220 (0.108) & 0.159 (0.082) & 0.295 (0.113) & 0.450 (0.141) & 0.316 (0.112) & 0.364 (0.123) & 0.789 (0.410) & 0.412 (0.132)\\
\bottomrule
\end{tabular}
\end{table*}

\paratitle{Within Group Agreement.} There are 2 groups of annotators (\ie MTurk workers and medical experts).
In this section, we study pairwise annotation agreement \textit{within each annotator group}.

As discussed in Section~\ref{sec:dataset}, there are 5000 documents annotated by MTurk workers and among them 200 documents are additionally annotated by experts. We report agreement within MTurk workers on the 5000 documents (MTurk-5000), and within medical experts on the 200 documents (Expert-200). Since these 200 documents also contain MTurk annotations, we also compute the agreement within MTurk workers on the 200 documents (MTurk-200), for a direct comparison with Expert-200.

\begin{observation}
The overall exact span agreement, within MTurk worker group and also within medical expert group, is very low.
\end{observation}

Table~\ref{tab:exact_and_relaxed_span_agree} reports averaged $F1$ values (standard deviations) of pairwise agreement within MTurk workers, and within experts, by different agreement evaluations.   
The overall exact span agreement within both groups is very low.
Among MTurk workers, the annotation agreement is extremely low. For labels I and O, $F1$ values are even lower than 0.1.
The similar low agreement is also observed among the expert annotators.
The agreement for label I is slightly higher than 0.5 but for labels P and O, $F1$ values are lower than 0.4.
These values show that more than a half of annotations are not agreed with other annotators, even among domain experts.
We believe that the low exact span agreement is caused by the high verbosity of PICO elements in medical literature.
As shown in Figure~\ref{fig:diff_anns_ex}, none of pairs of three annotations are agreed by the exact agreement, even though they indicate reasonable PICO information.

\begin{observation}
Annotators agree with the general areas where PIO elements appear, even though they made different choices in the start and end boundaries of annotations.
\end{observation}

On relaxed span agreement, both MTurk workers and medical experts show largely increased $F1$ than that of exact span agreement.
Specifically, on one-side boundary (OB) agreement, for MTurk worker group, $F1$ values are greater than twice of the $F1$ values estimated on exact span agreement. 
On token overlap (TO) agreement, for expert group, $F1$ values for all PIO elements are greater than 0.7, which is a clear indication of high level of agreement.
The improved agreement in relaxed span agreement demonstrates annotators made annotations at the similar areas but not necessarily with same boundaries.
For instance, in Figure~\ref{fig:diff_anns_ex} some pairs of annotations are agreed depending on the OB or TO agreements, and it is different from the zero agreed pair on the exact span agreement.

To summarize, due to the verbose descriptions of PICO elements in medical literature, annotations made by human annotators have very diverse boundaries. 
As the exact span agreement requires matching boundaries as well as labels, the low level of agreement is estimated for PICO span annotations.
However, the relaxed span agreements take into account the characteristics of PICO elements and show high level of agreement, by allowing the spans to have unmatched boundaries.

\paratitle{Cross Group Agreement.}
The finding, the low agreement in exact span agreement and the high agreement in relaxed span agreement, is observed between annotators who share similar understanding about domain knowledge (\ie MTurk-MTurk or Expert-Expert pairs).
In this section, we study the annotation agreement when annotators have different levels of domain knowledge, by measuring annotation agreement between MTurk workers and medical experts (\ie MTurk-Expert).  
Furthermore, we study differences in annotations between by MTurk workers and by medical experts in terms of lengths of span annotations.

On the 200 documents having annotations by both MTurk workers and medical experts, we measure the cross-group agreement. Specifically, we consider annotations by medical experts as gold standard annotations, and then estimate Precision, Recall, and $F1$ values for annotations by MTurk workers as predicted annotations. As each document has annotations from multiple MTurk and medical expert annotators, values derived by all possible MTurk-Expert pairs are averaged for a document.

Table~\ref{tab:inter_group_agreement} presents averaged Precision/Recall/F1 values of cross-group agreement on the 200 documents, with the three agreement types. In Table~\ref{tab:inter_group_agreement}, the cross-group agreements between MTurk workers and medical experts present the similar trend shown in within-group agreement (Table~\ref{tab:exact_and_relaxed_span_agree}).
The exact agreement is low and the relaxed agreements are much higher.
Based on the results of cross-group agreement as well as within group agreement, we make the third observation as follows.
\begin{observation}
Our finding, the high agreement on the general areas of PICO annotations with unmatched boundaries, is consistent regardless of domain knowledge that annotators have.
\end{observation}

\begin{figure*}[t]
  \includegraphics[width=0.75\linewidth]{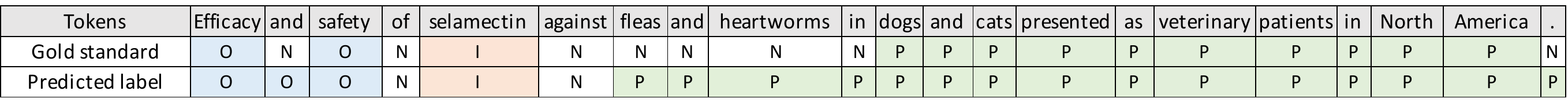}%
\caption{Examples of correct PICO spans and predicted PICO spans in a PICO span recognition task. The example sentence is from a medical document PMID: 10940525 in EBM-PICO dataset. N indicates a token without PICO label.}
\label{fig:hyper_representative_window}
\end{figure*}
Besides, an interesting finding is that when the agreement changes to relaxed from exact, Recall increases greater than 0.7 in the TO agreement while the improvement in precision is relatively small. This result shows that the annotations by MTurk workers are a `superset' of the annotations by medical experts in relaxed span agreements (high recall and low precision). Another finding is that on each agreement definition, the agreement in MTurk-Expert pair is always higher than the agreement in MTurk-MTurk pair, and lower than the agreement in Expert-Expert pair, comparing $F1$ values reported in both Tables~\ref{tab:exact_and_relaxed_span_agree} and~\ref{tab:inter_group_agreement}. It shows that some MTurk workers are capable of annotating PICO elements similar to medical experts.

\begin{table}
\centering
\caption{Comparison of the length of span annotations. The average (standard deviation in parenthesis) number of tokens are presented within MTurk workers on 5000 documents (MTurk-5000), and on 200 documents (MTurk-200) and medical experts on 200 documents (Expert-200).}
\label{tab:span_lengths}
\begin{tabular}{c|r@{}r|r@{}r|r@{}r}
\toprule
\textbf{Label} & \multicolumn{2}{c|}{\textbf{MTurk-5000}}   & \multicolumn{2}{c|}{\textbf{MTurk-200}}  & \multicolumn{2}{c}{\textbf{Expert-200}} \\
\midrule
P   & 9.355 &(10.406) &  8.268  &(9.113)  & 6.356  &(6.500) \\
I   & 4.356  &(7.999)  &  3.322  &(5.743)  & 1.903 & (2.075) \\
O   & 6.792  &(16.257) &  6.134  & ~(14.822) & 4.379 & (5.347) \\
\bottomrule
\end{tabular}
\end{table}

Next, we study differences in annotations made by the two groups. in terms of the number of tokens in each span annotation.
Table~\ref{tab:span_lengths} shows the average number of tokens and its standard deviation in each label of spans for the 5000 documents by MTurk workers, and for the 200 documents by MTurk workers and by medical experts.
Observe that the number of tokens in annotations by medical experts is smaller than the number of tokens in annotations by MTurk workers.
We believe the difference in the length of spans is attributed to medical domain knowledge.
Medical experts make more specific and brief annotations for being able to identify essential information.
Moreover, the annotations by MTurk workers show higher standard deviations than by experts.

\section{PICO Span Recognition}
\label{sec:PICORecognition}

Our evaluation shows that the exact span agreement is low and the relaxed span agreements are much higher, among human annotators, regardless within experts, within non-experts, or cross-group. In this section, we demonstrate how the exact and relaxed span agreements can be used for the evaluation of PICO span recognition task.
We also show how differently the exact and relaxed span agreements present the quality of performance.

The original EBM-PICO dataset paper also conducted a PICO recognition task. The performance is reported by token-level evaluation~\cite{PICO_dataset}.
That is, boundaries are not evaluated since they are predetermined as a token separation, and also the evaluation instances in token-level evaluation are more than that of span evaluation.
In this work, we conduct the same task as~\cite{PICO_dataset}, but evaluate the performance on span evaluation, specifically, the exact, OB, and TO span agreements.

For PICO span recognition, Bi-directional LSTM-CRF (BiLSTM-CRF) model~\cite{BLSTM_CRF} is used.
We follow the same experiment settings and train/validation/test data splits of the aggregated annotations (see section~\ref{sec:dataset}) used in \cite{PICO_dataset}.\footnote{\scriptsize 
From data exploration, it is found that about 10 percent of tokens have multiple labels in the aggregated annotations.
Before training a model, we resolve the multiple labels of tokens into a single label with the priority order of I, P, and O. We believe I element is the most important element among the three elements, followed by P and O.}  Table~\ref{tab:perf_PICO_recog} presents performance evaluated by the exact span agreement, and the OB and TO relaxed span agreements. As expected, low performance on the exact span agreement is observed, as it is challenging even for human annotators.
Performance evaluated by the two relaxed span agreements shows significant improvement. The values are even comparable to the results reported in Tables~\ref{tab:exact_and_relaxed_span_agree} and~\ref{tab:inter_group_agreement}.

Figure~\ref{fig:hyper_representative_window} shows examples of correct spans and predicted spans.
There are three predicted spans (\ie O, I and P in the order).
Indeed, the model correctly makes predictions on the general areas of correct spans.
However, boundaries of the predicted spans are incorrect except I element prediction.
Hence, by the exact span agreement, the only I predicted span is a correct prediction and the other two are incorrect predictions.
On the other hand, in the OB and TO agreements, 2 and 3 out of the predicted spans are considered as correct predictions, respectively.

\begin{table}
\footnotesize
\centering
\caption{Performance of BiLSTM-CRF in PICO span recognition evaluated by the exact and relaxed agreements. Precision/Recall/F1 values are estimated for each element label (P, I, O) and Micro-averaged value (Micro) for all labels.}
\label{tab:perf_PICO_recog}
\begin{tabular}{c|ccc|ccc|ccc}
\toprule
\textbf{Eval}& \multicolumn{3}{c|}{\textbf{Exact span}} & \multicolumn{3}{c|}{\textbf{One-side Boundary}}& \multicolumn{3}{c}{\textbf{Token Overlap}} \\
\midrule
Label& Pre & Rec  & $F1$ & Pre & Rec  & $F1$& Pre & Rec  & $F1$ \\
\midrule
P     & 0.227 & 0.205 & 0.216& 0.766 & 0.692& 0.727 & 0.840 & 0.758& 0.797\\
I     & 0.465  & 0.283& 0.352 & 0.792 & 0.481& 0.599& 0.835 & 0.508& 0.632\\
O     & 0.406 & 0.276 & 0.329 & 0.790 & 0.538 & 0.640& 0.838& 0.571 & 0.679 \\
\midrule
Micro   & 0.387 & 0.267 & 0.316& 0.785 & 0.541& 0.640& 0.837 & 0.577& 0.683\\
\bottomrule
\end{tabular}
\end{table}

\section{Conclusion}
\label{sec:Conclusion}

We report observations made from agreements in PICO span annotations.
The exact span agreement presents very low-level of agreement but the two relaxed span agreements show high-level of agreements in human annotations.
The result shows that even though boundaries of PICO annotations are unmatched, the
annotations are in the similar areas in general.
Based on our observations, we argue that the evaluation of PICO span-related tasks shall consider not only the exact span agreement but also the relaxed span agreements, because even human annotators do not agree on exact spans due to the high verbosity of PICO elements.

\bibliographystyle{ACM-Reference-Format}

\end{document}